\documentclass[aps,prb,twocolumn,floatfix,10pt,superscriptaddress]{revtex4-2}
\usepackage[colorlinks=true,bookmarks=false]{hyperref}
\usepackage{amsmath}
\usepackage{amssymb}
\usepackage{graphicx}
\usepackage{bm}
\usepackage{color}
\usepackage{upgreek}
\usepackage{scalerel}	
\usepackage{pgffor}
\usepackage{pdfpages}
\usepackage{float}
\usepackage{physics} 
\usepackage{lscape}   
\usepackage{comment}  
\usepackage{enumitem}

\usepackage{silence}
\makeatletter
\AtBeginDocument{\let\LS@rot\@undefined}
\makeatother

\newcommand{\moire}{moir\'{e}}
\newcommand\WeakJ{0.43t} 
\newcommand\WeakJSCF{0.42t} 
\newcommand\StrongJ{1.0t}

\newcommand{\BoldVec}[1]{ \ensuremath{ {\boldsymbol{#1}} } }

\usepackage[dvipsnames]{xcolor}

\newcommand{\TC}{T_{\rm c}}
\newcommand{\JC}{J_{\rm c}}
\newcommand{\TF}{T_{\rm F}}
\newcommand{\LM}{L_{\rm M}}
\newcommand{\KB}{k_{\rm B}}
\newcommand{\AC}{a_{\rm c}}
\newcommand{\NC}{N_{\rm c}}
\newcommand{\NP}{N_{\rm p}}
\newcommand{\KS}{k_{\rm s}}
\newcommand{\NK}{N_{\rm k}}
\newcommand{\Trans}{^{\rm T}}
\newcommand{\AAReg}{{\rm AA}}
\newcommand{\ABReg}{{\rm AB}}
\newcommand{\BAReg}{{\rm BA}}

\makeindex
\begin{document}

\title{%
	Nematic superconductivity in magic-angle twisted bilayer graphene from atomistic modeling 
}
\author{Tomas L\"{o}thman} 
	\email[Corresponding authors: ]{tomas.lothman@physics.uu.se }
	\affiliation{Department of Physics and Astronomy, Uppsala University, Box 516, S-751 20 Uppsala, Sweden}
\author{Johann Schmidt}
	\affiliation{Department of Physics and Astronomy, Uppsala University, Box 516, S-751 20 Uppsala, Sweden}
\author{Fariborz Parhizgar}
	\affiliation{Department of Physics and Astronomy, Uppsala University, Box 516, S-751 20 Uppsala, Sweden}
\author{Annica M. Black-Schaffer} 
	\email[]{annica.black-schaffer@physics.uu.se}
	\affiliation{Department of Physics and Astronomy, Uppsala University, Box 516, S-751 20 Uppsala, Sweden}

\date{\today}
\begin{abstract} 
	\begin{center}
	\textbf{Abstract} 
	\end{center}
	Twisted bilayer graphene (TBG) develops large \moire{} patterns at small twist angles with flat energy bands hosting domes of superconductivity. The large system size and intricate band structure have however hampered investigations into the superconducting state. Here, using full-scale atomistic modelling with local electronic interactions, we find at and above experimentally relevant temperatures a highly inhomogeneous superconducting state with nematic ordering on both atomic and \moire{} length scales. The nematic state has a locally anisotropic real-valued $d$-wave pairing, with a nematic vector winding throughout the \moire{} pattern, and is three-fold degenerate. Although $d$-wave symmetric, the superconducting state has a full energy gap, which we tie to a $\pi$-phase interlayer coupling. The superconducting nematicity is further directly detectable in the local density of states. Our results show that atomistic modeling is essential and also that very similar local interactions produce very different superconducting states in TBG and the high-temperature cuprate superconductors.
\end{abstract}

\maketitle
\section{Introduction}
Precise twist angle and carrier density control have made it possible to map the rich phase diagram of twisted bilayer graphene (TBG) \cite{Cao2018, Cao2018a}. Around the magic twist angle $\theta \approx 1.1^\circ$ four spin-degenerate topological \moire{} flat bands are formed around zero energy and the associated density of states (DOS) peaks make TBG extremely prone to electronic ordering \cite{Li2009,Po2018,Cao2018, Cao2018a, Lu2019, Kerelsky2019, jiang2019charge, Xie2019, Choi2019,Sharpe2019,Balents2020, andrei2020graphene}. Both correlated insulating states at integer flat band fillings ($\nu=\pm 1,2,3$) and superconductivity have been found, with the latter appearing as superconducting domes both flanking the $\nu = \pm 2$ insulating states \cite{Cao2018a,Yankowitz2019} and more broadly across the entire \moire{} bands \cite{Lu2019}. 

While the superconducting transition temperature $\TC \approx 3$~K is low in TBG \cite{Cao2018a,Lu2019}, the ratio to the Fermi temperature is large $ \TC / \TF \approx 0.1$, exceeding the weak coupling regime \cite{andrei2020graphene}. In fact, with superconductivity appearing in close proximity to correlated insulators and a pseudogap state with reduced DOS above the superconducting domes \cite{jiang2019charge,Xie2019,Choi2019} with accompanied strange metal behavior \cite{Polshyn2019, Cao2020}, the phase diagram of TBG shares striking similarities to the high-temperature cuprate and pnictide superconductors \cite{keimer2015quantum, chubukov2012pairing, andrei2020graphene}. This points to the possibility of strong local electronic interactions being responsible for superconductivity, although electron-phonon pairing is also a plausible candidate \cite{Peltonen2018,Wu2018,Polshyn2019,Lian2019}.

The large scale of the \moire{} pattern ($\sim10^4$ carbon atoms at the magic angle) and the non-trivial band topology have severely hampered studies of superconductivity in TBG. Continuum models reproduce the normal-state band structure \cite{Bistritzer2011,Moon2013}, but are harder to reconcile with the possibility of strong local electronic interactions. Effective lattice models, using e.g.~the \moire{} pattern or otherwise rescaled lattice schemes, have been used in Hubbard-like tight-binding studies, but must strike a difficult balance between accuracy and orbital proliferation \cite{Su2018}. Many of these studies have proposed a time-reversal symmetry (TRS) breaking chiral $d$-wave state \cite{Su2018,PhysRevB.103.L041103,PhysRevB.98.085436, PhysRevLett.121.087001, PhysRevB.99.195114, PhysRevB.98.241407, PhysRevLett.121.217001, PhysRevB.97.235453}, as also found in heavily doped graphene \cite{BlackSchaffer2007, Nandkishore2012}.

Finding chiral $d$-wave superconductivity is at first sight not surprising, considering that the $d_{x^2-y^2}$-wave state of the cuprates naturally transforms into	the two $d_{x^2-y^2}$- and $d_{xy}$-wave states with a symmetry enforced degeneracy at $\TC$, on lattices with three- or six-fold symmetry, such as graphene's honeycomb lattice and the TBG \moire{} lattice \cite{Nandkishore2012, BlackSchaffer2014, Venderbos2018, Chichinadze2020}. On the honeycomb lattice, the chiral $d_{x^2-y^2}+id_{xy}$-wave combination then becomes fully gapped, while any real-valued nematic $d$-wave state always has a nodal spectrum, thus making the chiral state energetically favored and the ground state at zero temperature \cite{BlackSchaffer2007, Nandkishore2012,BlackSchaffer2014}. A similar favoring could naively also be expected to hold in TBG. However, in apparent contrast to these basic expectations, different nematic superconducting states have very recently been proposed for TBG. Some of these require higher order coupling to additional coexisting orders or pairing channels to energetically favor the nematic state over the chiral state \cite{PhysRevB.99.144507, Dodaro2018, Chichinadze2020, Wang2021}. More intriguing are direct findings, i.e.~without other orders or pairing channels, of a nematic phase in a phase diagram region using either rescaled or atomistic lattice models, although with seemingly varying properties \cite{Su2018, Julku_2020, PhysRevB.103.L041103}. Experimentally, recent magnetotransport measurements have also identified intrinsic nematicity in the superconducting state of TBG \cite{Cao2021}. Taken together, these results all highlight the question of if and how nematic superconductivity appears in TBG, as well as its measurable consequences.

In this work, to accurately capture superconductivity in TBG, we use full-scale atomistic modeling, including all carbon atoms, a dense $k$-point sampling, and electron interactions in agreement with both current TBG experiments and known interactions in graphene(-like) systems, as well as consistent with the cuprates. By solving both the mean-field self-consistent and linear gap equations using realistic transition temperatures, we find a highly inhomogeneous and real-valued nematic $d$-wave superconducting state, with both atomic- and \moire-scale nematicity, dominating the phase diagram and at the experimentally observed critical temperatures. We further show directly measurable signatures of the nematicity in the local density of states. Unexpectedly, the nematic $d$-wave state also has a full energy gap, which we tie to a strong $\pi$-phase Josephson locking between the graphene layers.		

\section{Results}
\subsection{Normal state properties}
			\begin{figure*}
					\centering
						\includegraphics[width=0.98\textwidth]{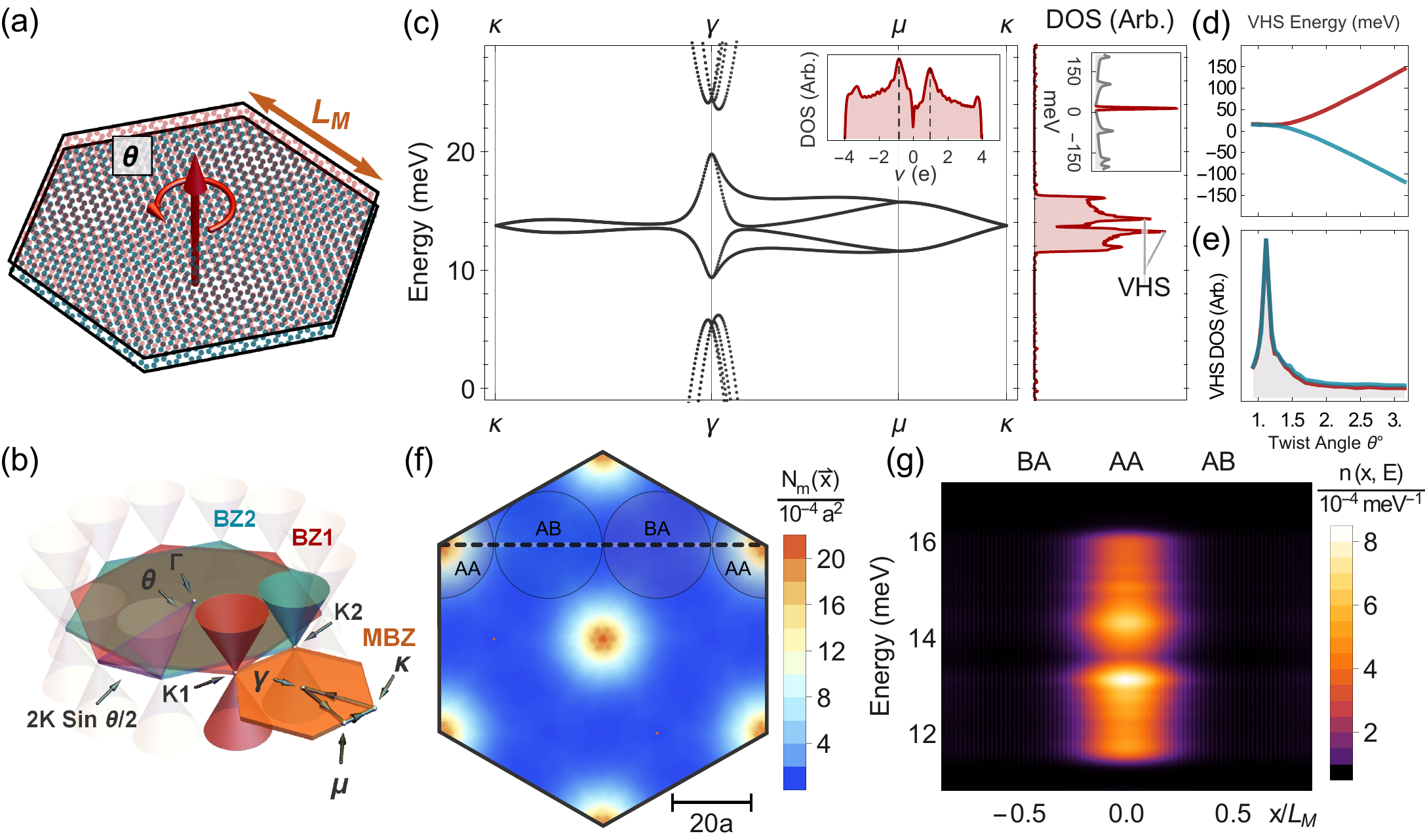}
					\caption{%
						{\bf Lattice structure and normal state properties of twisted bilayer graphene (TBG).} 
						{\bf (a)} Carbon lattice in one \moire{} cell of TBG with twist angle $\theta \approx 3.9^\circ$ and \moire{} period $\LM$.
						{\bf (b)} Illustration of TBG in reciprocal space, where $\theta$ rotates the two hexagonal Brillouin zones (BZ) of the  graphene layers (BZ1 and BZ2), displacing the Dirac cones at the BZ corners (${\rm K1}$ and ${\rm K2}$) by $\Delta K = 2K\sin(\theta/2)$, the reciprocal length of the \moire{}{} Brillouin zone (MBZ). Depicted are also the high symmetry lines and points ($\gamma$, $\mu$, $\kappa$) in the MBZ.
						{\bf (c)} 
						Low energy band structure of TBG at $\theta\approx1.2^\circ$ along the high symmetry lines in (b), consisting of four narrow spin-degenerate \moire{} bands (solid lines) separated from other bands (dotted lines) by finite energy gaps. The corresponding large density of states (DOS), peaking at the upper and lower van Hove singularities (VHSs), is plotted on same energy axis to the right. Side inset: DOS for a larger energy range, highlighting the massive \moire{} flat band (red) DOS peak. Inset: DOS as a function of the \moire{} band filling factor $\nu$ in units of electrons per \moire{} unit cell with VHSs at $\nu \approx \pm 1$ (dashed lines).
						{\bf (d)} Energy of the upper and lower VHSs as a function of twist angle $\theta$.
						{\bf (e)} Maximum peak height of the two VHS peaks as a function twist angle $\theta$, peaking sharply at the magic angle.
						{\bf (f)} Intensity plot of the top layer charge density $N_{\rm m}(\bm{x})$ of the completely filled \moire{} bands, equivalent to integrating the LDOS in (g). The scale bar shows a distance of $20a$, where $a=2.46$~\AA\ is the graphene lattice constant.						 
						{\bf (g)} Local density of states (LDOS) $n (\bm{x},E)$ as a function of energy $E$ and position along the dashed line in (f), centered on the $\AAReg$ region.
					}
					\label{fig:01 - Normal}
			\end{figure*}
		TBG consists of two graphene layers rotated by a twist angle $\theta$, which produces a periodic \moire{} interference pattern of length $\LM = a/(2\sin{(\theta/2)})$ \cite{Morell2010,Bistritzer2011}, far exceeding the graphene lattice constant $a=2.46$~\AA, see Fig.\,\ref{fig:01 - Normal}(a). We model commensurate angle TBG with a full-scale tight-binding model of all carbon atoms in the large periodic \moire{} unit cell. The out-of-plane $p_z$ carbon orbitals hybridize in-plane through standard nearest neighbor hopping $t=2.7$~eV, while the interlayer hybridization is captured by an exponentially decaying hopping together with a Koster-Slater angular dependence \cite{Shallcross2010,Moon2013}, see Methods. 
				
		In reciprocal space, the Dirac cones from the graphene layers are displaced by the twist and sit at the corners of the smaller \moire{} Brillouin zone (MBZ), see Fig.\,\ref{fig:01 - Normal}(b). As the twist angle decreases, the Fermi velocity of the Dirac cones is reduced. The layer and valley degrees of freedom then form four spin-degenerate narrow bandwidth \moire{} bands that separate from the remaining band structure \cite{Bistritzer2011}, see Fig.\,\ref{fig:01 - Normal}(c). As a consequence, TBG has a large DOS around zero energy, peaking in two van-Hove singularities (VHSs) that correspond to saddle points in the band structure \cite{Bistritzer2011,Li2009}. At the magic angle, the \moire{} bands become essentially completely flat and the two VHSs merge, see Figs.\,\ref{fig:01 - Normal}(d,e). 
		
		The states of the \moire{} bands are primarily localized to the $\AAReg$ region of the \moire{} cell, where the carbon atoms of the two layers are aligned, as shown by the inhomogeneous and three-fold symmetric charge density $N_{\rm m}(\bm{x})$ in Fig.\,\ref{fig:01 - Normal}(f). Resolving the charge density with respect to energy, Fig.\,\ref{fig:01 - Normal}(g) further shows the local density of states (LDOS) along a periodic path passing the $\ABReg$, $\AAReg$, and $\BAReg$ regions (black dotted line in Fig.\,\ref{fig:01 - Normal}(f)).
		
\subsection{Modeling superconductivity}
Many properties of the superconducting state in magic-angle TBG are still unknown. We therefore employ a general model for the superconducting pairing, guided by only a few constraints: the weak interlayer van der Waals coupling motivates intralayer pairing, while the observed suppression of superconductivity in in-plane magnetic fields and no evidence for spin-polarized Cooper pairs, restrict us to consider spin-singlet pairing \cite{Cao2018a,Wu2019}. The strong on-site repulsion in graphene, graphite \cite{Wehling2011}, and TBG \cite{Kerelsky2019} also points to spin-singlet pairing, since in the strong coupling limit of the Hubbard model, the resulting $t$-$J$ model gives spin-singlet nearest-neighbor bond interactions \,\cite{Lee_2006,BlackSchaffer2007}. In fact, preference for bond spin-singlet configurations, over double and single occupancies, was already central in early treatments of $p\pi$-bonded planar organic molecules (of which graphene is the infinite extension) \cite{Pauling1960}, and subsequently also used for cuprate superconductors \cite{Anderson1987}. Spin-singlet bond pairing has also recently been derived from spin-fluctuations in TBG \cite{PhysRevB.103.L041103} and also used in smaller rescaled lattice models \cite{Su2018,Julku_2020}. Consequently, we model superconducting pairing in TBG by spin-singlet order parameters on every in-plane nearest-neighbor carbon bond, using a uniform coupling strength $J$, see Methods. Together these bond order parameters forms an order parameter field $\bm{\hat{\Delta}}$ throughout the \moire{} cell.
		
We solve for the superconducting order parameter field with each bond order parameter treated fully independently using both the full non-linear, self-consistent and the linearized gap equations, see Methods. The linear gap equation is valid at $\TC$ and has the same symmetry as the normal state, leading to its solutions always belonging to one of the irreducible representations of the symmetry point group $D_3$ of TBG: the one-dimensional $A_1$ and $A_2$, or the two-dimensional $E$ representation. Below $\TC$, non-linear contributions enter, possibly further breaking the symmetry, which we capture by iteratively solving the full non-linear and self-consistent gap equation at zero temperature. This combined approach enables us to completely characterize the superconducting state.
		
			\begin{figure*}[htb!]
				\centering
					\includegraphics[width=1.0\textwidth]{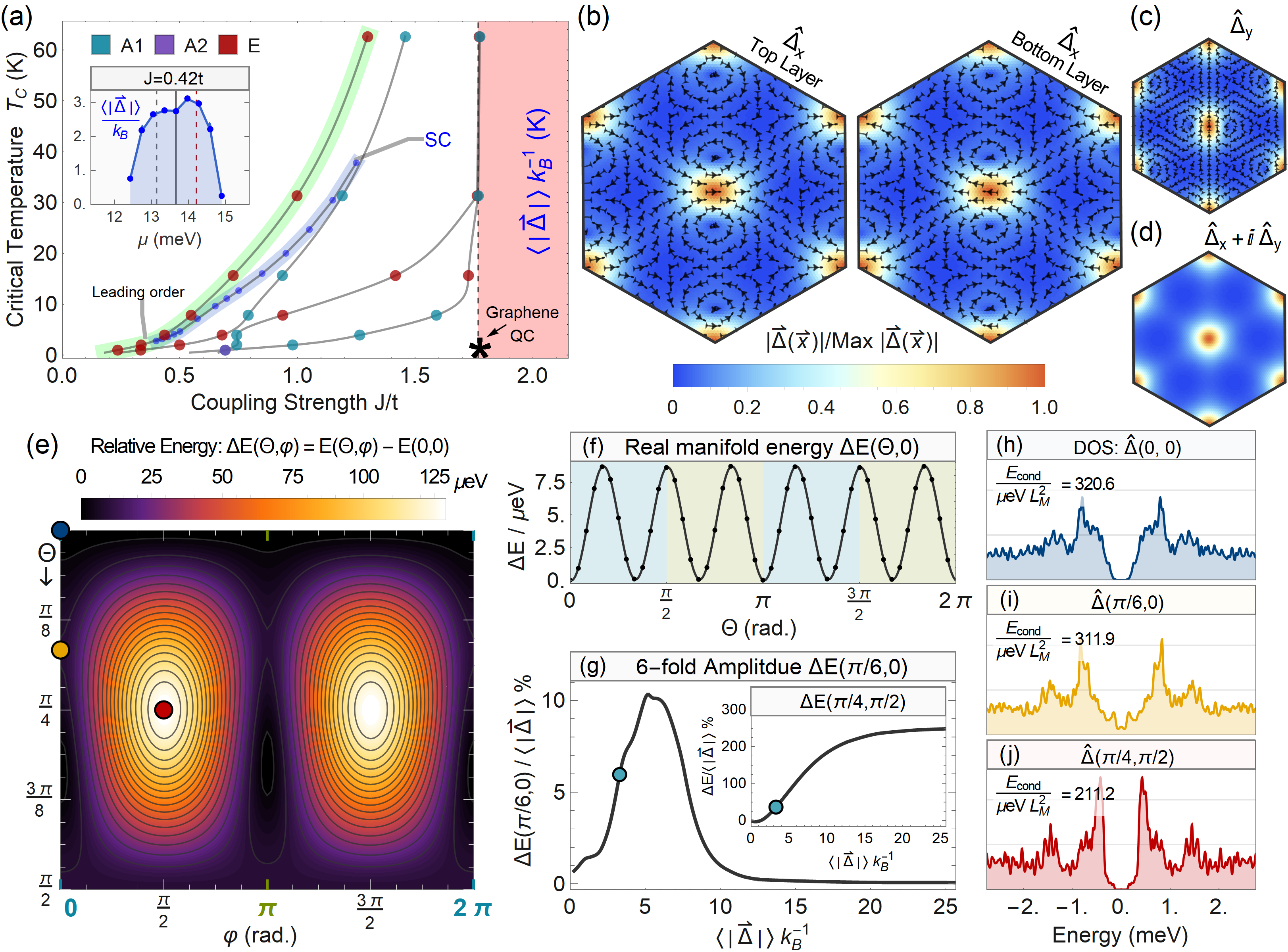}
				\caption{%
					{\bf Properties of the nematic superconducting state in twisted bilayer graphene (TBG).}
					{\bf (a)} Critical temperatures $\TC$ (left $y$-axis) obtained from the linear gap equation as a function of the coupling strength $J/t$ with the Fermi level at the upper van Hove singularity (VHS). Plot marker color labels the symmetry of each solution, gray lines guide the eye. The leading two-fold degenerate solution is highlighted in green. Highlighted in blue are the spatial averaged amplitude order parameter, $\langle |\bm{\Delta}(\bm{x}_i)| \rangle \KB^{-1}$, (right $y$-axis)
					obtained from the non-linear self-consistency (SC) equation at zero temperature ($T=10^{-7}t$).
					Inset: Self-consistent order parameter as a function of chemical potential $\mu$ at weak coupling $J = \WeakJSCF$. 
					Vertical lines mark upper VHS (dashed red), lower VHS (dashed black), and half-filling (solid black).
					{\bf (b)} Plot of the order parameter field $\bm{\hat{\Delta}_x}$ throughout the \moire{} cell in the top (left) and bottom graphene layer (right). Underlying color intensity plot shows the amplitude, clearly depicting a global, \moire-scale, nematicity. The streamlines of the local vectors field $\bm{\chi}(\bm{x})$ illustrate the intricate atomic-scale nematicity in the local $d$-wave bond order parameters, as well as the superconducting $\pi$-phase-shift between the two layers.
					{\bf (c)} Same as (b) but for $\bm{\hat{\Delta}}_y$ in the top graphene layer.
					{\bf (d)} Color intensity plot of the order parameter field amplitude for the chiral $\bm{\hat{\Delta}}_x + i \bm{\hat{\Delta}}_y$ solution in the top graphene layer.
					{\bf (e)} Relative free energy landscape at $T=0$ for all independent linear combinations $\bm{\hat{\Delta}}(\Theta,\varphi)$ of the $\bm{\hat{\Delta}}_{x,y}$ solutions.
					{\bf (f)} Cut along the real-valued linear combinations of (e), which due to gauge invariance are located at the lines $\varphi=0,2\pi$ (blue) and $\varphi=\pi$ (green) in (e). The line is a least square fit, showing a three-fold degenerate nematic ground state. 
					{\bf (g)} Energy difference between the most and least favorable nematic combinations in (f) as a function of tuning the averaged order parameter amplitude, with dot marking value for $J=\WeakJ$. Inset: Energy difference between most favorable nematic state and the time reversal symmetry (TRS) breaking chiral $\bm{\hat{\Delta}}_x + i \bm{\hat{\Delta}}_y$ combination.
					{\bf (h)}, {\bf (i)}, {\bf (j)} Low-energy density of states (DOS) for most favorable nematic, least favorable nematic, and chiral combination, respectively, extracted at the points of the same color in (e). Plots show that the ground state is fully gapped, while the other combinations show a (partially) filled energy gap reflecting lower superconducting condensation energies (given in figures). For all but panels (a) and (g): $J=\WeakJ$ and chemical potential aligning the Fermi level with the upper VHS (band filling $\nu \approx 1$), but conclusions are unchanged for other $J$. 
				}
				\label{fig:02 - TceV}
			\end{figure*}	
			
\subsection{ Moir\'e-scale nematicity}	
		Starting with the results from the linear gap equation, we show in Fig.\,\ref{fig:02 - TceV}(a) the four highest critical temperatures $\TC$ as a function of coupling strength $J$ at the magic angle and with the Fermi level aligned with the upper VHS, corresponding to $\nu \approx 1 $. Across the wide range of all investigated coupling strengths, the solution with highest $\TC$ (green highlight) belongs to the $E$ irreducible representation and is therefore two-fold degenerate and spanned by two real-valued order parameter fields: $\bm{\hat{\Delta}}_x$ and $\bm{\hat{\Delta}}_y$. In Fig.\,\ref{fig:02 - TceV}(b) we plot in color scale the normalized amplitude of the $\bm{\hat{\Delta}}_x$ order parameter field in the upper (left) and lower (right) graphene layer in the \moire{} cell, while the amplitude of $\bm{\hat{\Delta}}_y$ in the upper layer is plotted in Fig.\,\ref{fig:02 - TceV}(c). Both $\bm{\hat{\Delta}}_x$ and $\bm{\hat{\Delta}}_y$ clearly break the $C_3$ rotation symmetry of the normal state and are instead enhanced in the $\AAReg$ regions and along the $C_2$ nematic axes aligned with the $x$- and $y$-axis, respectively. 
				
		In contrast to pristine graphene where superconductivity only develops for $J/t$ above the quantum critical point (QCP) $\JC=1.76t$ \cite{Uchoa2007,BlackSchaffer2007,Kopnin2008}, TBG shows substantially enhanced ordering in Fig.\,\ref{fig:02 - TceV}(a), with finite a $\TC$ even at very weak coupling. For weak coupling, $\TC$ is approximately proportional to $J/t$, as expected for an isolated flat band \cite{Heikkila2011, Kopnin2011, Volovik2013, Lothman2017}. For slightly stronger coupling, the growth in $\TC$ however accelerates, especially when approaching the graphene QCP. This both indicates contributions from dispersive bands \cite{Julku_2020} and suggests that the low energy \moire{} flat bands have a catalytic effect that at least partially triggers the QCP cascade boosting $\TC$ at couplings also below the QCP.
				
		Alongside $\TC$ extracted from the linear gap equation, we also plot in Fig.\,\ref{fig:02 - TceV}(a) (blue line, right $y$-axis) the spatially averaged amplitude of the self-consistent order parameter field, obtained from the full non-linear gap equation at zero temperature (in Kelvin). As seen, $\TC$ and the averaged order amplitude are directly comparable, despite the strong amplitude inhomogeneity over the \moire{} unit cell. After fixing $J$ to match the experimentally observed $\TC \approx 4$~K \cite{Cao2018a,Lu2019}, we tune the chemical potential across all \moire{} band fillings. The resulting zero temperature self-consistent order amplitude is shown in the inset of Fig.\,\ref{fig:02 - TceV}(a). The superconducting amplitude predictably drops sharply at the two \moire{} band edges. In between, the amplitude is rather stable but has two local maxima near the upper and lower VHS (dashed lines) and a local minima at half filling (solid black line), largely replicating the features of the normal-state DOS in Fig.\,\ref{fig:01 - Normal}(c).
			
		Further analyzing the zero temperature self-consistent solutions at different experimentally relevant couplings and band fillings, we always find that they are exclusively real-valued, even when self-consistency is achieved iteratively from an initial complex number field configuration. In fact, we find that the self-consistent solutions are always real-valued linear combinations of the two leading and degenerate solutions $\bm{\hat{\Delta}}_{x}$ and $\bm{\hat{\Delta}}_{y}$ of the linear gap equation, see Methods. This makes TBG a nematic superconductor\,\cite{Fu2014}, with nematic symmetry breaking on the \moire{} length scale. As both the normal-state and electronic interactions are fully isotropic, this nematicity is entirely due to spontaneous symmetry breaking in the superconducting state.
							
\subsection{Degenerate ground state.}
	To further investigate the ground state symmetry at zero temperature, we introduce the parametrization
	$
		\bm{\hat{\Delta}} (\Theta,\varphi) = 
		\| \bm{\hat{\Delta}} \| 
		\left( 
			\cos \Theta \bm{\hat{\Delta}}_x 
			+ e^{i \varphi} \sin \Theta \bm{\hat{\Delta}}_y 
		\right)
	$.
	Here, $\| \bm{\hat{\Delta}} \|$ is the norm of the order parameter field, the nematic angle $\Theta$ controls the angle between the $C_2$ nematic axis and the $x$-axis, and a relative complex phase, $\varphi\neq 0$, breaks TRS. Here, the $(\Theta,\varphi) \in [0,\pi/2] \times [0,2\pi]$ manifold is a (Bloch) sphere because the $\Theta = 0, \pi/2$ lines collapse to points due to gauge invariance.
	
	At $\TC$ all the states $\bm{\hat{\Delta}}(\Theta,\varphi)$ are degenerate solutions, but due to non-linear contributions of a finite order field only specific combinations are expected to be energetically favorable at zero temperature. Demonstrating this symmetry lifting, we show in Fig.\,\ref{fig:02 - TceV}(e) the relative free energy of all possible $\bm{\hat{\Delta}}(\Theta,\varphi)$ at zero temperature, confirming the self-consistent results, the free energy minima are achieved for the real-valued linear combinations. The free energy maxima (i.e.~smallest condensation energy) are instead attained around the rotationally symmetric (see Fig.\,\ref{fig:02 - TceV}(e)) TRS breaking complex combinations $\bm{\hat{\Delta}}(\pi/4, \pm\pi/2) = \bm{\hat{\Delta}}_x \pm i \bm{\hat{\Delta}}_y$. The free energy maxima (i.e.~smallest condensation energy) are instead attained around the rotationally symmetric (see Fig.\,\ref{fig:02 - TceV}(e)) TRS breaking complex combinations $\bm{\hat{\Delta}}_\pm = \bm{\hat{\Delta}}(\pi/4, \pm\pi/2) = \bm{\hat{\Delta}}_x \pm i \bm{\hat{\Delta}}_y$.

	On closer inspection, the energy splitting among the real-valued combinations is six-fold symmetric as a function of the nematic angle $\Theta$, see Fig.\,\ref{fig:02 - TceV}(f). As a result, $\bm{\hat{\Delta}}_x$ is one of the three gauge-inequivalent nematic ground states, all with the nematic $C_2$-axis directed towards one of the next-nearest-neighbor $\AAReg$ regions.  In contrast, $\bm{\hat{\Delta}}_y$ is one of the three least stable gauge-inequivalent nematic states that instead has the $C_2$-axis directed towards one of the nearest-neighbor $\AAReg$ regions, as seen in Fig.\,\ref{fig:02 - TceV}(b,c).			
		
	The amplitude of the energy splitting among the real-valued nematic states depends non-monotonically on the field norm (or equivalently on $J$ or $\TC$), as seen in Fig.\,\ref{fig:02 - TceV}(g), with notably large values, up to $10\%$ of $\KB \TC$, around the experimentally observed $\TC$. 
		In contrast, the energy of the TRS breaking chiral state $\bm{\hat{\Delta}}_\pm$ relative to the nematic ground state is large around the experimentally observed critical temperatures and also increases monotonically with the field norm (or $\TC$ or $J/t$). We thus find that the nematic state is heavily favored in this regime, and even more so at stronger coupling. Based on these results it is also not surprising that nematic states have  previously also been reported in the strong coupling regime \cite{Su2018, PhysRevB.103.L041103}. While any degeneracy lifting of the $E$ manifold must necessarily vanish as the field norm goes to zero, we do find that the chiral state, $\bm{\hat{\Delta}}_{\pm}$, just barely becomes the ground state (beyond the resolution used in Fig.~\ref{fig:02 - TceV}(g)) for very small field amplitudes, before transitioning to the nematic state for more realistic temperatures. 	
																			
	The energy split among the different superconducting states is further reflected in the different gap structures among the $\bm{\hat{\Delta}}(\Theta,\varphi)$ states, as revealed by the low energy DOS in Figs.\,\ref{fig:02 - TceV}(h)-(j) at an experimentally relevant temperature. The ground state $\bm{\hat{\Delta}}_x = \bm{\hat{\Delta}}(0,0)$ is fully gapped with a resulting large condensation energy achieved by pushing occupied VHS states down in energy in to sharp coherence peaks. In contrast, the least favored nematic states, including $\bm{\hat{\Delta}}_y$, are not fully gapped, in turn limiting the condensation energy. For the least favored of all the states in the $E$ manifold, the TRS breaking chiral state $\bm{\hat{\Delta}}_{\pm}$, we find a substantially reduced gap with coherence peaks that are closer together, both features substantially reducing the resulting condensation energy. 
								
	Based on the results above, we have demonstrated that TBG is a nematic superconductor at experimentally relevant temperatures with a strong dependence on the orientation of the $C_2$ nematic axis affecting both the condensation energy and gap structure. In particular, the real-valued nematic ground state is energetically favored due to a fully gapped spectrum, which is in sharp contrast to superconductivity in doped graphene where the nematic state is always nodal while the chiral $d$-wave state is energetically favored and has a full energy gap \cite{BlackSchaffer2007, Nandkishore2012, Kiesel2012}. In fact, our findings of a $d$-wave nematic state with a full energy gap is overall unexpected as $d$-wave symmetry traditionally is associated with a nodal energy spectrum. However, it has recently been shown that the situation can be more complicated in multiorbital or multiband systems, where completely nodeless and gapped $d$-wave superconducting states has recently been found \cite{Chubukov2016, Agterberg2017, Pang2018, Nakayama2018, Nica2021}. This points towards the large scale \moire{} pattern and the multiple nearly degenerate \moire{} bands as having an important role in determining the properties of the superconducting state. In particular, we find that the gap structure sensitively depends on the fine distinctions that arise with different $C_2$ nematic axis orientations, and we also note that the gap structure of previously reported nematic $d$-wave states have been found to be nodal in rescaled lattice and continuum model while unreported in recent atomistic model \cite{Su2018,PhysRevB.99.195114,PhysRevB.103.L041103}. Moreover, we note that our results show that the real-valued nematic ground state is favored over the complex chiral combination directly in the $E$ pairing channel of a simple pairing model. Our nematic ground state is therefore produced without any higher order coupling to additional coexisting orders or nearly degenerate pairing channels that have additionally been shown to otherwise favor nematicity \cite{PhysRevB.99.144507, Dodaro2018, Chichinadze2020}. 
				
\subsection{Atomic-scale \texorpdfstring{$d$}{d}-wave nematicity}	
	Having established a nematic superconducting state in TBG on the \moire{} length scale, we next characterize the superconducting state on the atomic scale, where the symmetry is well described by the point group $D_{6h}$ of the graphene honeycomb lattice. We accomplish this by introducing a three-dimensional local order vector $\bm{\Delta}(\bm{x})$ for the three nearest-neighbor bond order parameters of each carbon and projecting this vector onto the complete set of form factors $\bm{f}$ for the $D_{6h}$ irreducible representations: an extended $s$-wave and two $d$-waves, $d_{xy}$ and $d_{x^2-y^2}$, see Methods. Because the nematic state has a pure local $d$-wave character with a negligible $s$-wave component ($<0.025\%$) and is also completely real, each local order vector is uniquely expressed as a real-valued linear combination:		
		$
			\bm{\Delta}(\bm{x}_i) =
			|\bm{\Delta}(\bm{x}_i)| 
			\left(
				\cos \tau(\bm{x}_i) \bm{f}_{d_{x^2-y^2}} + \sin \tau(\bm{x}_i) \bm{f}_{d_{xy}}
			\right)
		$,
	where the vector field $\bm{\chi}(\bm{x}_i) = \cos \tau(\bm{x}_i) \bm{\hat{x}}+ \sin \tau(\bm{x}_i) \bm{\hat{y}}$ captures the spatially varying $d$-wave orientation. As shown by the streamlines of the vector field $\bm{\chi}(\bm{x}_i)$ in Figs.\,\ref{fig:02 - TceV}(b,c), there exists a strong atomic-scale variation in the nematicity, which is aligned with the \moire-scale nematic axis in the central $\AAReg$-region, but then also forms two vortex-antivortex pairs of opposite circulation outside the $\AAReg$ region. The two antivortices are always pinned to the center of the $\ABReg$ and $\BAReg$ regions, independently of the $C_2$ nematic axis orientation, while the two vortices stay close to the $\AAReg$-region on opposite sides of the nematic axis, following its orientation. This atomic-scale nematic pairing vortex pattern consistently appears for all investigated filling factors, coupling strengths $J/t$, and in both the zero temperature self-consistent linear and the non-linear gap equation solutions, making it a very robust feature of the nematic state.

	We also note that this nematic pairing vortex pattern is distinct from superficially similar patterns of spontaneous supercurrents found in TBG with a chiral ground state \cite{Su2018,PhysRevB.99.195114,PhysRevB.103.L041103}, since the time-reversal invariant nematic ground state is necessarily free from supercurrents. Still, a vortex pattern has also been found in the normal-state Dirac mass term in a TBG continuum model \cite{Zhang2019}, which together might suggest an overall tendency for vortex formation within large \moire{} patterns.
	
\subsection{Nematic signatures in LDOS}	
			\begin{figure*}[ht]
				\centering
					\includegraphics[width=1.0\textwidth]{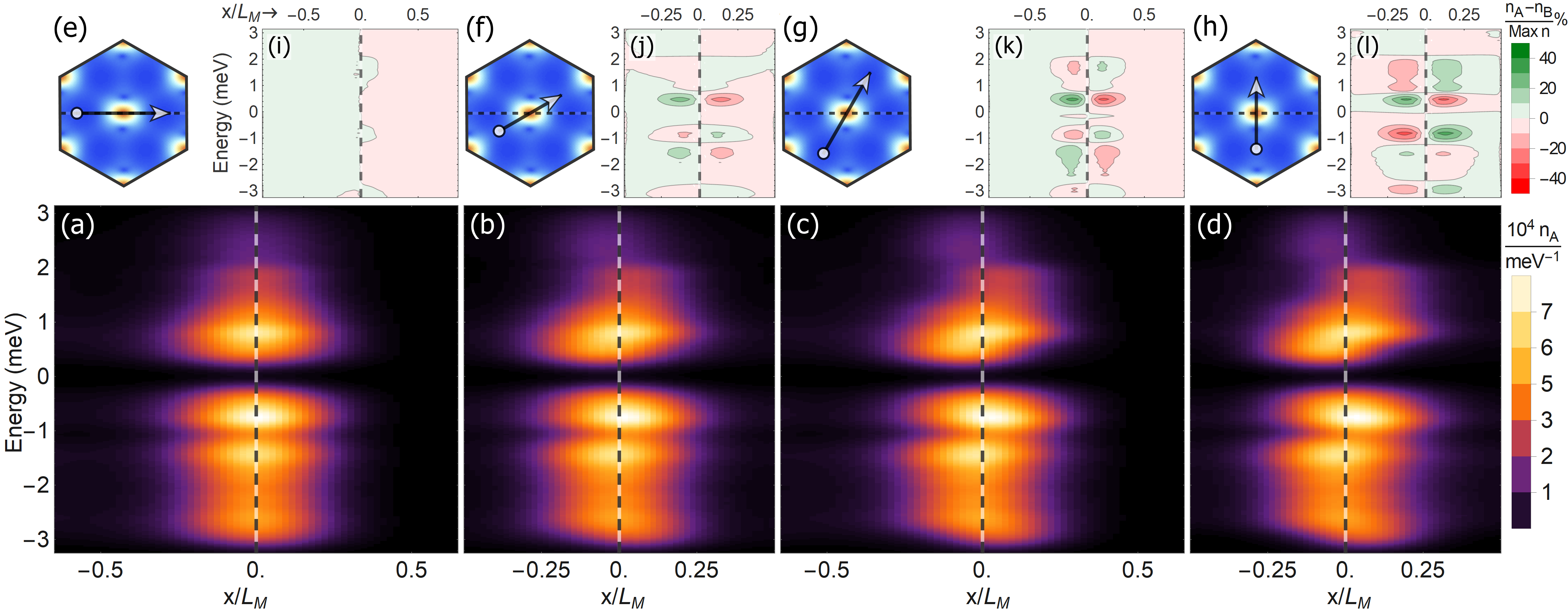}
				\caption{%
					{\bf Nematic signatures in the local density of states (LDOS).} 								
					{\bf (a)-(d)} Sublattice resolved (sublattice $A$) top layer electronic LDOS $n_{\rm A}$ of the nematic superconducting state contour plotted along four different real space cuts (black line shown in (e)-(h)). The \moire{} cell center, $x / \LM = 0$, is marked by dashed lines. 
					{\bf (e)-(h)} Illustrations of the line cuts along which the LDOS is shown in (a)-(d), starting at the base and ending at the arrow tip. The cuts are shown on top of the superconducting order parameter field amplitude where the dashed line shows the nematic axis of the order. In the normal state, the cuts (a) and (c) are equivalent by symmetry, and so are cuts (b) and (d). The normal state LDOS is therefore the same in each pair of cuts and also symmetric around the \moire{} cell center, but this symmetry is lifted by the nematic superconducting order in {(a)-(d)}. 					
					{\bf (i)-(l)} Contour plots of the sublattice LDOS polarization, i.e~the LDOS difference between sublattice $A$ and $B$, $n_{\rm A}-n_{\rm B}$, along the cuts of {(e)-(h)}, where green shades indicate a larger LDOS on sublattice $A$, while red shades indicate a larger LDOS on sublattice $B$. The LDOS polarization is measured relative to the maximal LDOS, $ \text{Max} \: n$, and each contour marks a 10\% change relative to this maximal LDOS. In all figures, the coupling strength is $J=\WeakJ$ and the Fermi level is aligned with the upper van Hove singularity (conclusions are qualitatively unchanged for other $J/t$).
				}
				\label{fig:03 - SCSTM}
			\end{figure*}
		
		The rotational symmetry breaking of the nematic superconducting state is clearly visible in the superconducting order parameter: the \moire-scale nematic $C_2$ axis, the intricate atomic-scale $d$-wave nematicity pattern, and the three-fold degenerate ground state. Nematic superconductivity is also known to exhibit magnetic field directional dependencies in various properties tied to superconductivity, such as the upper critical field, as also recently found in superconducting TBG \cite{Cao2021}. 
		
		In Fig.\,\ref{fig:03 - SCSTM}, we show that the nematic superconducting state in TBG additionally gives rise to measurable signals in the electronic local density of states (LDOS). In the main panels, (a) to (d), we plot the LDOS on sublattice $A$ in the top layer along the corresponding four different line cuts shown in (e) to (h). In the normal state, the three-fold rotational symmetry leads to equivalent LDOS on the cuts of (a),(c) and (b),(d), respectively. In the superconducting state, however, the four cuts in (e) to (h) form an progressively increasing angle against the nematic $C_2$ axis (dashed line), and the corresponding LDOS in (a) to (d) show an increasingly pronounced asymmetry around the \moire{} cell center, unambiguously demonstrating the nematic superconducting order. In fact, the nematic superconducting state induces shifts, primarily in the coherence peaks, that are up to almost half the maximal \moire{} state LDOS in size. The sublattice LDOS polarization, i.e. the difference in LDOS between the $A$ and $B$ sublattices, is shown in (i) to (l) along the cuts, demonstrating that all shifts occur in opposite directions between the two sublattices. This means the observed anisotropy washes out in a joint LDOS, which further highlights the importance of atomistic modeling and might also explain why no nematic anisotropy in the LDOS has been found in the superconducting state in rescaled lattice models \cite{Su2018}. To summarize, our results show that the nematic superconducting state in TBG is clearly observable in sublattice-resolved scanning tunneling spectroscopy/microscopy (STS/STM) measurements.

\subsection{Interlayer \texorpdfstring{$\pi$}{π}-locked Josephson coupling}	
			\begin{figure}[htb]
				\raggedright
					\includegraphics[width=0.46\textwidth]{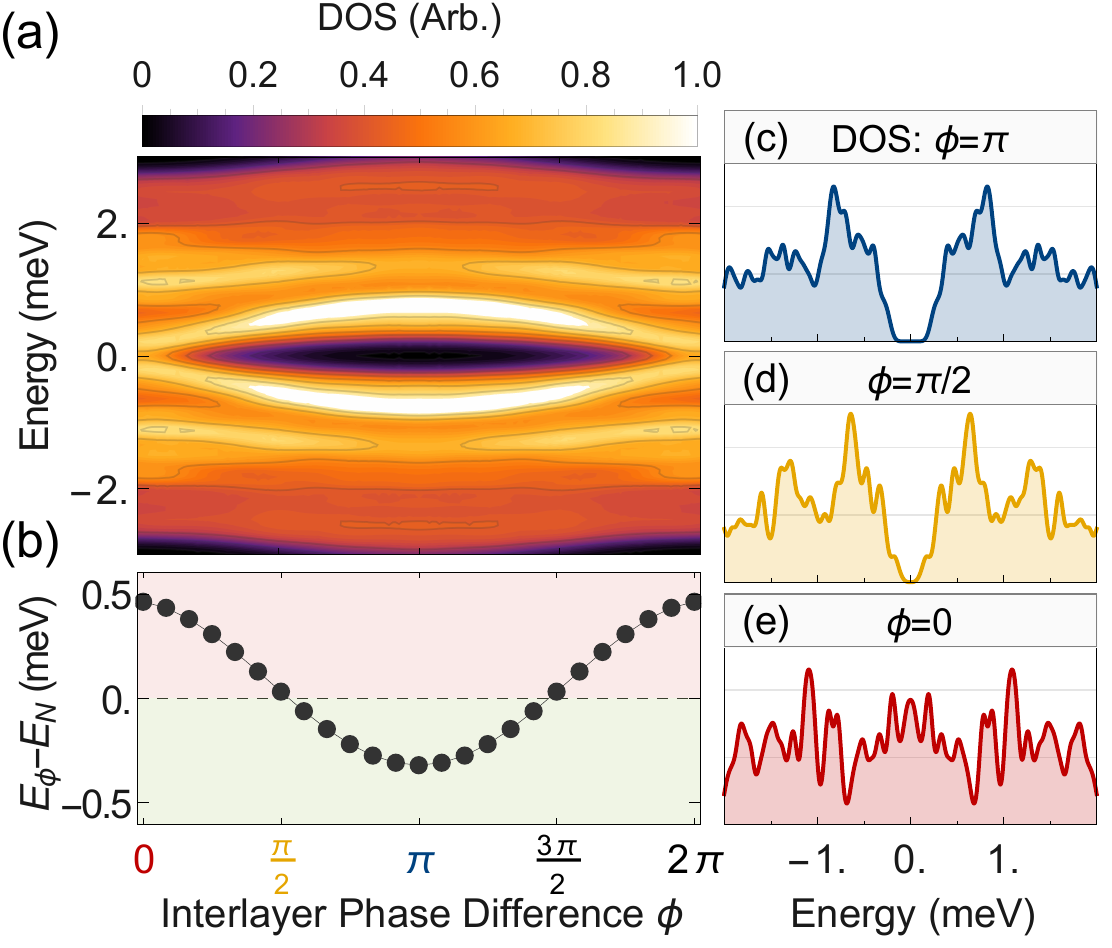}
				\caption{
					{\bf Interlayer Josephson coupling.}
					{\bf (a)} Color intensity plot of the density of states (DOS) as a function of interlayer phase difference $\phi$, tuned by shifting the phase of the order parameter in the bottom layer starting from the the self-consistent solution found at $\phi=\pi$. Away from $\phi=\pi$  energy gap and coherence peaks are monotonically reduced with the gap closing around $\phi=0,2\pi$.
					{\bf (b)} Superconducting condensation energy, i.e.~energy of superconducting state relative to the normal state (N) as a function of interlayer phase $\phi$. Pink region marks positive condensation energy, describing an unstable superconducting state.
					{\bf (c)-(e)} DOS for interlayer phases $\phi=0, \pi/2, \pi$, respectively, taken as vertical cuts in (a). Here, the coupling constant is $J=\WeakJ$ and Fermi level at the upper van Hove singularity (conclusions are qualitatively unchanged for other $J$).
				}
				\label{fig:04 - JC}
			\end{figure}
		The superconducting order parameters at both $\TC$ and zero temperature always show a rigid interlayer $\pi$-shift in the superconducting phase and the atomic-scale nematic vectors in Fig.\,\ref{fig:02 - TceV}(b) consequently reverse direction between the two layers. This result is consistent with earlier continuum model results predicting a $\pi$-shift due to a large layer counterflow velocity \cite{Bistritzer2011,PhysRevB.99.195114}, and was also recently reported numerically using atomistic modeling \cite{PhysRevB.103.L041103}, albeit seemingly missing for a rescaled model \cite{Su2018}.	This rigid $\pi$-shift suggests a strong and unconventional interlayer Josephson coupling that we here explore by manually tuning the interlayer phase of the self-consistent order parameter, as is standard in Josephson setups.
		
		Adding a phase factor $\bm{\Delta} \mapsto e^{i \phi} \bm{\Delta}$ in the bottom layer, we plot in Fig.~\ref{fig:04 - JC}(b) the resulting superconducting condensation energy as a function of the interlayer phase difference $\phi$. For roughly half of the available phase space, the superconducting state is unstable with a total energy higher than in the normal state. This extraordinarily stiff energy-phase relationship proves that the $\pi$-shift is central for the existence of superconductivity.			 
		
		The underlying reason for the strong interlayer Josephson coupling is that the $\pi$-shift is responsible for the entire superconducting gap of the nematic $d$-wave superconducting state of TBG, as seen in the DOS color intensity plot in Fig.~\ref{fig:04 - JC}(a), with specific spectra extracted in Figs.~\ref{fig:04 - JC}(c-e). At the self-consistent solution, $\phi=\pi$, a full superconducting gap is present, whereas it decreases and eventually disappears for $\phi$ approaching zero (or equivalently $2\pi$). This result is in notable contrast to any real-valued $d$-wave state in graphene or $\AAReg$ stacked bilayer graphene, which are always nodal, regardless of interlayer phase difference. We thus conclude that the \moire{} band structure generates the interlayer $\pi$-shift, which in turn results in the nematic $d$-wave state of TBG being fully gapped and the ground state solution.

\section{Conclusion}
	Using full-scale atomistic modeling of magic-angle TBG, we find a highly inhomogeneous and nematic $d$-wave superconducting state with a full energy gap at experimentally observed critical temperatures. The large real-space inhomogeneity demonstrates that atomic-scale modeling is crucial for TBG. Besides atomic spatial resolution, we also find that a dense $k$-point sampling of the MBZ and the \moire{} flat band structure is required to correctly capture the superconducting order. In fact, if sampling only the MBZ center, we instead find a chiral $d$-wave ground state \cite{PhysRevB.103.L041103} at the experimentally relevant temperatures. 
	
	The only possible uncertainty left in our work is the exact nature of the electronic interactions, where we assume intralayer spin-singlet bond interactions. This choice is well motivated both by experimental evidence \cite{Cao2018a,Wu2019,Kerelsky2019} and theoretical work on TBG \cite{PhysRevB.103.L041103}, graphene \cite{BlackSchaffer2007, Wehling2011, Kiesel2012}, and $p\pi$-bonded organic molecules \cite{Pauling1960}. Similar interactions are also present in the cuprate superconductors \cite{Lee_2006}, with which TBG shares many phase diagram features \cite{andrei2020graphene}. Moreover, even if additional longer-range interactions were considered, results from the honeycomb lattice indicate that our results will remain qualitatively correct \cite{Nandkishore2012, Kiesel2012}. 

	The only unknown parameter is then the coupling constant $J/t$. Our results are however remarkably stable with the same nematic ground state for all experimentally relevant coupling strengths $J/t$, further supporting their validity. Quantitatively, we also find experimentally observed $\TC$:s already at very weak coupling $J/t\sim0.4$, which clearly illustrates the strong tendency of the \moire{} flat bands to electronically order \cite{Cao2018a,Cao2021}. Thus, the required coupling strength $J/t$ is easily attained and also compares favorably with order of magnitude estimates of the super-exchange $J/t = 4 t/U \sim1$ from graphene based on ab-initio results and a strong coupling scenario \cite{Wehling2011,Kerelsky2019}.

	We also note that we calculate the superconducting  mean-field pair amplitude $\Delta$, while superconductivity in two dimensions is reached only at the Berezinskii-Kosterlitz-Thouless (BKT) transition, marking the onset of phase coherence and requiring a finite superfluid weight. Here our use of the full TBG band structure with its non-trivial topology guarantees a finite geometric superfluid weight \cite{Peotta2015,Hu2019,Julku_2020,Xie2020a}. While the BKT temperature is always substantially lower than the mean-field temperature, they have been shown to correlate directly  in several TBG models \cite{Julku_2020}. Thus, the mean-field pair amplitude calculated in this work should be a good measure of superconductivity, albeit it requires us to consider appreciably higher mean-field critical temperatures than those measured experimentally.

	Finally, let us comment on the correlated insulating states existing at integer fillings \cite{Li2009, Po2018, Cao2018, Cao2018a, Lu2019, Kerelsky2019, jiang2019charge, Xie2019, Choi2019,Sharpe2019,Balents2020, andrei2020graphene}. In addition to the superconducting state, a complete phase diagram should be supplemented by these interspersing insulating states. With the insulating states surviving to higher temperatures, superconductivity then primarily appears as domes flanking the insulating regions when tuning the filling. In fact, this is a universal feature of flat band systems: superconductivity always thrives further away from high DOS peaks compared to any particle-hole (insulating) order, thus naturally forming domes \cite{Lothman2017}. This universal picture is consistent with current TBG experiments and notably independent of the mechanisms behind the insulating behavior and superconductivity being the same or competing in nature. As such, our results are not sensitive to the exact nature of the insulating states. In conjunction with this, it is interesting to point out that, despite similar phase diagram features and electronic interactions for the cuprate superconductors and TBG \cite{andrei2020graphene}, we find a remarkably different (fully gapped, highly inhomogeneous, and nematic) $d$-wave superconducting state in TBG. We might expect any insulating state to also display similar atomic-scale inhomogeneity, opening up for a wide range of different behaviors in TBG and other \moire{} systems.			

\section{Methods} \label{sec:Method}
\footnotesize

\noindent {\bf Moir\'{e} lattice structure.} 
\label{sec:Moire}
	In the first layer of TBG, the two graphene unit vectors are  
		$\bm{a}_{11} = a \bm{\hat{x}}$, 
		$\bm{a}_{12} = (a/2)\bm{\hat{x}} + (3a/2) \bm{\hat{y}}$,
	while in the second layer, the lattice is rotated by the relative twist angle $\theta$
	and spanned by the rotated unit vectors $\bm{a}_{2i} = R(\theta) \bm{a}_{1i}$, 
	where $R(\theta)$ is the rotation matrix in two dimensions. 
	With two carbon atoms per graphene unit cell, 
	there are in both layers two sublattices, $A$ and $B$, of carbon atoms placed at 
		$
			\{ 
				n \bm{a}_{l1} + m \bm{a}_{l2} 
				+ \delta_{S B} \bm{\eta}_l + (d_0 \delta_{l} / 2) \bm{\hat{z}}
			\}
	$
	for $n, m \in \mathcal{Z}$, $l=1,2$, and $S \in A,B$, where 
		$\bm{\eta}_1 = \AC \bm{\hat{y}}$, $\bm{\eta}_2 = R(\theta) \bm{\eta}_1$, $\AC = a/\sqrt{3}$, 
		$\bm{\hat{z}}$ is the unit vector perpendicular to the layers, 
		$d_0 = 3.35$~\AA{}, 
		and $\delta_l=(-1)^l$.
	The two layers produces a periodic \moire{} interference pattern of period length 
		$\LM = a / (2\sin{\theta/2})$.

	We perform calculations using a large and periodically repeated \moire{} unit cell with the unrelaxed lattice positions \cite{Wijk2015,Nam2017,Lucignano2019}. 
	This requires using commensurate twist angles, for which the lattices of the two layers periodically match up: 
		$n_1 \bm{a}_{11} + n_2 \bm{a}_{12} = m_1 \bm{a}_{21} + m_2 \bm{a}_{22}$, 
	for integers $m_i, n_i$. 
	Twist angles satisfying this condition are given by  
		$\cos \left( \theta \right) = (3q^2-p^2)/(3q^2+p^2)$ 
	and parametrized by a relative prime integer pair ($\gcd(p,q)=1$, greatest common divisor), 
	with $p,q \in \mathbb{N}$, $p \ge q > 0$ \cite{Shallcross2010}. 
	Specifically, the number of carbon atoms in the \moire{} unit cell is	
	$N_{\rm c} (p,q) = 4 \gcd (3,p) \left(p^2 + 3q^2\right) / \gcd (p-3q,p+3q)^2 $. 
	We choose $p=1$ with $q$ odd, as this results in the fewest number of carbon atoms within the \moire{} unit cell for a given twist angle. 
	In particular, we focus in the main text on the commensurate unit cell $(p,q)=(1,55)$, which gives the twist angle $\theta \approx 1.2^\circ$ closest to the magic angle, defined as where the Fermi velocity vanishes. 
	We here also note that the commensurate condition is independent of the twist center. Since we use a $\gcd (3,p)=1$ commensurate \moire{} lattice constructed from an $\AAReg$ stacked bilayer and twisted around an axis passing through two carbon atoms, the resulting lattice has a $D_3$ point symmetry group. Due to the long wave length \moire{} pattern at small twist angles, this lattice also has an an approximate $D_6$ symmetry that even becomes exact if the twist axis is instead taken through the center of a honeycomb lattice hexagon \cite{Zou2018}.

\noindent {\bf Normal state Hamiltonian.} 
\label{sec:TB}
	We employ a standard tight-binding model including all carbon atoms:
		$H_{\rm N}  =  H_{\rm intra} + H_{\rm inter}$.
	The intralayer Hamiltonian is given by graphene:
		\begin{align}
		\label{eq:intrahopping}
			H_{\rm intra} &=  
				-  \mu \sum_{i, \sigma, l}  c_{i \sigma l}^{\dagger} c_{i \sigma l}
				- t \sum_{\substack{ \langle i, j \rangle \\ \sigma, l}}  
					\left( 
						c_{i \sigma l}^{\dagger} c_{ j\sigma l} + \text {H.c.} 
					\right),
		\end{align} 
	with in-plane nearest neighbor hopping $t$ between the carbon $p_z$ electrons, created by the operator $c_{i,\sigma,l}^{\dagger}$ on the site $i$, layer $l$, and spin $\sigma$. The overall occupancy is regulated by the chemical potential $\mu$. Further, the interlayer hybridization has been shown to be well-captured by hopping amplitudes decaying exponentially with distance and with Koster-Slater factors for the bond angle dependence\,\cite{Moon2013}:
	\begin{align}
		\label{eq:interhopping}
			H_{\rm inter} &= 
			- \sum_{i, j} t_{ij} \left( c_{i \sigma 1}^{\dagger} c_{j \sigma 2} + \text {H.c.} \right)
			\\
			t_{ij} &= 
			-t e^{(\AC - r_{ij})/ \lambda} \sin^2 (\beta^z_{ij})  + 
			t_{\sigma} e^{(d_0-r_{ij})/ \lambda} \cos^2 \beta^z_{ij},
	\end{align}
	where $\bm{r}_{ij}$ is the displacement between the carbon sites $i$ and $j$ and 
		$\cos \beta^{z}_{ij} = \bm{\hat{r}}_{ij} \cdot \bm{\hat{z}}$. 
	The input parameters are fixed by matching to the electronic band structure single and $\ABReg$-stacked bilayer graphene \cite{Moon2013}: $t = 2.7\,\mathrm{eV}$, $t_{\sigma} = 0.48\,\mathrm{eV}$, and $\lambda=0.184 \AC$. To keep the Hamiltonian matrix sparse, we cutoff $t_{ij}$ for distances $d > 6 a$, resulting in $\sim 250$ interlayer bonds per atom. This long-ranged cutoff is needed to preserve all symmetries of the TBG lattice, while a larger cutoff does not affect our results.
	
\noindent {\bf Superconducting pairing.} \label{sec:SCmethod}
	We model superconductivity by:
	\begin{align}
			\label{eq:SCModel}
			H_{\rm SC} 
			= 
			\sum_{ \langle i,j \rangle } 
				\Delta_{ij}
				\left( 
					c^\dagger_{i\uparrow} c^\dagger_{j\downarrow} - 
					c^\dagger_{i\downarrow} c^\dagger_{j\uparrow} 
				\right) 
				+ {\rm H.c.} 
				+ \frac{|\Delta_{ij}|^2}{J},
		\end{align}
		with a homogenous coupling strength $J$. We solve for the superconducting spin-singlet bond order parameters $\Delta_{ij}$ using both the full non-linear self-consistent and the linearized gap equations:
		\begin{gather}
					\Delta_{ij} =  - J \langle s_{ij} \rangle \label{eq:selfcons}
					\\
					\delta\Delta_{ij} 
						= \sum_{ \langle r, s \rangle } 
							\left[ 
								-J \frac{\partial \langle s_{ij} \rangle}{\partial \Delta_{rs} } 
							\right] 
							\delta\Delta_{rs} 
						= 
						\sum_{ \langle r, s \rangle } 
							S^{rs}_{ij}(T) \delta\Delta_{rs}, \label{eq:linearized}
		\end{gather}
		where
			$
				s^\dagger_{ij} = 
					c^\dagger_{i\uparrow} c^\dagger_{j\downarrow} - 
					c^\dagger_{i\downarrow} c^\dagger_{j\uparrow} 
			$. 
	The non-linear self-consistency gap equation, Eq.~\eqref{eq:selfcons}, is equivalent to $\partial F / \partial { \bar{\Delta}_{ij} } = 0$, and thus finds the order parameters $\Delta_{ij}$ which minimize the Free energy $F$. This equaqtion is solved iteratively until convergence is reached.	The linear gap equation, Eq.~\eqref{eq:linearized}, only has solutions for the infinitesimal order field $\delta\bm{\hat{\Delta}}$ (defined up to a complex constant) when the stability matrix $S^{rs}_{ij}(T)$ has at least one eigenvalue equal to $1$, which implicitly defines the critical temperature $\TC$. This equation is equivalent to the Free energy Hessian 
		$
			\partial^2 F /  
			\partial \bar{\Delta}_{i j} 
			\partial \Delta_{r s}
			= 
			\partial \langle s_{ij} \rangle / \partial \Delta_{rs}
			+
			\delta_{ir} \delta_{js} / J 
		$
		changing signature with a zero eigenvalue in the normal state ($\bm{\hat{\Delta}}=0$) and the emerging order therefore lowers the Free energy below $\TC$.
		
	\noindent {\bf Rational pole expansion.} 
		Solving the gap equations, Eqs.~\eqref{eq:selfcons}-\eqref{eq:linearized}, using the standard approach of matrix diagonalization is prohibitively expensive for TBG due to large degrees of freedom. We instead calculate the Fermi operator $F_{\beta}(H) = (e^{\beta H} +1)^{-1}$ using a rational pole expansion \cite{Goedecker1993,Goedecker1999}. Specifically, we use the minimax rational approximation of J.~E.~Moussa that minimizes the uniform norm $ \epsilon = \max_{z \in [-\beta E_{min}, \infty]}| F_{\beta}(z) - \sum_{n=1}^{\NP} { R_n / (\beta z - P_n) } | $ for $\NP$ poles at $P_n$ with residues $R_n$, because of its rapid convergence at low temperatures \cite{Moussa2016}.
		
		To illustrate the method, we first show that all single particle (anomalous) expectation values are elements of $F_{\beta}( H_{\text{BdG}} (\bm{k}) )$, within the Bogoliubov de-Gennes (BdG) formalism. In the $2\NC$ dimensional block Nambu-spinor basis 
		$ 
			X_{\bm{k}} = 
			(
				\{c_{ \bm{k}, \uparrow } \}, 
				\{c_{-\bm{k}, \downarrow	}^\dagger\}
			)\Trans 
		$, 
	the complete Hamiltonian $H = H_{\rm N} + H_{\rm SC}$ for the $\NC$ carbon atoms and spins has the BdG bilinear form	
	\begin{equation}
	\label{Eq:BdGForm}
		\begin{split}			
			H
			& =
			\sum_{ \bm{k} }
				X_{ \bm{k} }^\dagger
				\begin{pmatrix}
					H_{\rm N} (\bm{k})						& 	\Delta (\bm{k}) 				\\
					\Delta^\dagger (\bm{k}) 	& 	-H\Trans_{\rm N} (-\bm{k}) 	
				\end{pmatrix}
				X_{ \bm{k} }
				+	C
			\\
			&	=
			X_{\bm{k}}^\dagger	H_{\text{BdG}} (\bm{k}) X_{ \bm{k} }
			+	C
			\,,
		\end{split}
	\end{equation}		
	with the constant energy shift $C = - \mu \NC + \sum_{\langle i,j \rangle} |\Delta_{ij}|^2 / J$. The BdG bilinear form is diagonalized by a unitary transformation with the block structure		
		\begin{gather} 
			U (\bm{k})		
			=
			\begin{pmatrix}
				u	  (\bm{k})				&	v  	(\bm{k})	 \\
				\bar{v} (-\bm{k})	&	\bar{u} (-\bm{k})			
			\end{pmatrix}
			\\ 
			U (\bm{k})^\dagger H_{ \text{BdG} } (\bm{k}) U (\bm{k})				
			=	\mathcal{E}(\bm{k}) 
			= \text{diag}( \{ E_{\bm{k}} \}, \{ -E_{-\bm{k}} \} ).
		\end{gather}
	The accompanying canonical transformation defines the fermionic Bolgoliubov quasiparticles of opposite energy $\pm E$,
		\begin{equation}	
			\label{eq:BQP}
				Y_{\bm{k}} = 
				(
					\{ \gamma_{ \bm{k}, \uparrow		},        \}
					\{ \gamma_{-\bm{k}, \downarrow	}^\dagger \}
				)\Trans 			
				\qquad
				X_{\bm{k}} = U(\bm{k}) Y_{\bm{k}}.
		\end{equation}
	In this diagonal basis, the Fermi operator is
		\begin{equation}
			F_{\beta} ( H_{\text{BdG}} (\bm{k}) ) 
			=
			U(\bm{k}) F_{\beta} ( \mathcal{E}(\bm{k}) )  U^\dagger(\bm{k}) 
			=
			\begin{pmatrix}
				G^{<}  		(\bm{k})				& 	F 		(\bm{k})	\\
				F^\dagger (\bm{k})				&		G^{>} (\bm{k})		
			\end{pmatrix}
		\end{equation}
	where the blocks are:
		\begin{align*}
			G^{<} (\bm{k}) & =
				u(\bm{k}) 		F_{\beta}( E_{ \bm{k} } ) u^\dagger (\bm{k})	+ 
				v(\bm{k}) 		F_{\beta}(-E_{-\bm{k} } ) v^\dagger (\bm{k})
			\\                                   
			G^{>} (\bm{k}) & =                            
				\bar{v} (-\bm{k}) F_{\beta}( E_{ \bm{k} } ) v\Trans (-\bm{k}) + 
				\bar{u} (-\bm{k}) F_{\beta}(-E_{-\bm{k} } ) u\Trans (-\bm{k})
			\\                                   
			F (\bm{k}) & =                            
				u(\bm{k}) 		F_{\beta}( E_{ \bm{k} } ) v\Trans (-\bm{k}) + 
				v(\bm{k}) 		F_{\beta}(-E_{-\bm{k} } ) u\Trans (-\bm{k}) 
		\end{align*}
		From the transformations of Eq.~\eqref{eq:BQP}, we find that all single particle expectation values are indeed given by the elements of $F_{\beta} ( H_{\text{BdG}} (\bm{k}) )$:
		\begin{equation}
			\begin{split}
				\label{Eq:ExplicitAExp}
				\langle	c^\dagger_{\bm{k} i} c_{\bm{k} j} \rangle
				& =
				\sum_{\nu}
					\left[						
						u_{j \nu} (\bm{k})
						F_\beta ( E_{\bm{k} \nu} )
						u^{\dagger}_{\nu i} (\bm{k}) 
					\right.
				 \\ &
					\quad +
					\left.		
						v_{j \nu} (\bm{k})
						F_\beta ( -E_{-\bm{k} \nu} )
						v^{\dagger}_{\nu i} (\bm{k})				
					\right]
				\\
				&	=
					\BoldVec{e}^{\dagger}_j 
					F_{\beta}( H_{\text{BdG}} (\bm{k}) ) 
					\BoldVec{e}_i
					= [G^{<} (\bm{k})]_{ji}
			\end{split}
		\end{equation}		
		\begin{equation}
			\begin{split}
				\langle	c_{-\bm{k} i}	c_{\bm{k} j} \rangle
				& =
				\sum_{\nu}
					\left[ 			
						u_{j \nu} (\bm{k}) 
						F_\beta ( E_{\bm{k} \nu} )
						v\Trans_{\nu i} (-\bm{k}) 						
					\right.
					\\ &
					\quad +
					\left.
						v_{j \nu} (\bm{k}) 
						F_\beta ( -E_{-\bm{k} \nu} )
						u\Trans_{\nu i} (-\bm{k})						
					\right]
					\\
					& = 
					\BoldVec{e}^{\dagger}_j 
					F_{\beta} ( H_{\text{BdG}} (\bm{k}) ) 
					\BoldVec{h}_i
					= [F(\bm{k})]_{ji}
			\end{split}
		\end{equation}
		where we introduce $2\NC$ dimensional basis vectors for the electron $[\BoldVec{e}_i]_j = \delta_{ij} $ and hole $ [\BoldVec{h}_i]_j = \delta_{(i+\NC)j}$ blocks of the BdG Matrix, respectively. 
		
		Next, we apply the rational pole expansion. Specifically, the (anomalous) expectation values for $\BoldVec{q}_i=\BoldVec{e}_i$ ($\BoldVec{q}_i=\BoldVec{h}_i$) are
		\begin{align}
			\langle	c_{-\bm{k} i}^{(\dagger)}	c_{\bm{k} j} \rangle
			&  =
			\BoldVec{e}^{\dagger}_j 
			F_{\beta} ( H_{\text{BdG}} (\bm{k}) ) 
			\BoldVec{q}_i
			\nonumber \\ &
			\approx
			\sum_{n=1}^{\NP} { R_{n} \BoldVec{e}^{\dagger}_j \left[ \beta H_{\text{BdG}} (\bm{k}) - P_n \right]^{-1}  \BoldVec{q}_i }
			\nonumber \\ &
			=
			\sum_{n=1}^{\NP} { R_{n} \BoldVec{e}^{\dagger}_j \Pi^{n}_{ \bm{k} }  \BoldVec{q}_i }
			=
			\sum_{n=1}^{\NP} { R_{n} \BoldVec{e}^{\dagger}_j \BoldVec{p}^{n}_i },
			\label{eq:PFMEXP}
		\end{align}
		where we define the "propagated" vectors $\left[ \beta H_{\text{BdG}} (\bm{k}) - P_n \right]^{-1}  \BoldVec{q}_i = \Pi^n_{ \bm{k} } \BoldVec{q}_i = \BoldVec{p}^{n}_i$.
		Thus, by solving the set of linear equations for the propagated vectors
			\begin{equation}
				\label{eq:ExpLinearSystem}
					\left[ \beta H_{ \text{BdG} } ( \bm{k} )  - P_n \right]
					\BoldVec{p}^{n}_i 
					= 
					\BoldVec{q}_i, 			
			\end{equation}
		we can calculate the anomalous expectation values in Eq.~\eqref{eq:selfcons} within the rational pole expansion of the Fermi operator using $\NP$ terms.
		
		Finally, we also need access to derivatives, or generally a static response, in Eq.~\eqref{eq:linearized}. Given a perturbation $\lambda$ to $H_{\text{BdG}}$, the static response of any expectation value, can also be computed using the rational pole expansion. Namely, using that
		$
			\partial_{\lambda} A^{-1}_{\lambda} 
			=	 -A^{-1}_{\lambda}  [ \partial_{\lambda} A_{\lambda} ] A^{-1}_{\lambda} 
		$ for a matrix $A_{\lambda}$, the static response is,
		\begin{align}
			\frac{\partial \langle	c_{-\bm{k} i}^{(\dagger)}	c_{\bm{k} j} \rangle }{ \partial \lambda }
			\approx	
			- \sum_{n=1}^{\NP} { 
				\beta R_{n} 
				\BoldVec{e}^{\dagger}_j 
				\left[
					\Pi^{n}_{ \bm{k} }  
					\frac{ \partial H_{\text{BdG}} (\bm{k})  }{ \partial \lambda }
					\Pi^{n}_{ \bm{k} }
				\right]
				\BoldVec{q}_i
				.
			}
		\end{align}
		In particular, for derivatives with respect to the superconducting order parameter $\Delta_{rs}$, only the off-diagonal blocks $\bm{\hat{\Delta}}(\bm{k})$ are non-zero. In addition, $ H_{\text{BdG}} (\bm{k}) $ is block diagonal and contains only the normal state Hamiltonian when evaluated at $\bm{\hat{\Delta}}=0$, as is the case for the linear gap equation. As a consequence we find
		\begin{align}
			\frac{\partial \langle	c_{-\bm{k} i}	c_{\bm{k} j} \rangle }{ \partial \Delta_{rs} }
			\approx	
			\sum_{n=1}^{\NP} { 
				\beta R_{n}
				( \BoldVec{y}^{n}_j )^\dagger
				\left[
					\frac{ \partial \bm{\hat{\Delta}} (\bm{k})  }{ \partial \Delta_{rs} }
				\right]
				\BoldVec{x}^{n}_i 
			}
			\label{eq:PFMResp}
		\end{align}
		where $\BoldVec{x}^{n}_i$ and $\BoldVec{y}^{n}_j$ are $\NC$ dimensional vectors satisfying the linear equations: 
		\begin{equation}
		\label{eq:RespLinearSystem}
				\left[ \beta H_{\rm N} (\bm{k}) - P_n	\right] \BoldVec{x}^{n}_{i} = \BoldVec{i}
			\quad 
				\left[\beta \bar{H}_{\rm N}(-\bm{k}) + \bar{P}_n \right] \BoldVec{y}^{n}_{j}  = \BoldVec{j}
		\end{equation}
		where $[\BoldVec{i}]_s = \delta_{is}$ and  $[\BoldVec{j}]_s = \delta_{js}$.
		
		Together, Eq.~\eqref{eq:PFMEXP} and Eq.~\eqref{eq:PFMResp}, show how both (anomalous) expectation values and their static response are computed using the rational pole expansion. Thus solving the gap equations Eqs.~\eqref{eq:selfcons}-\eqref{eq:linearized} is simply reduced to solving the respective sets of linear equations of Eqs.~\eqref{eq:ExpLinearSystem}-\eqref{eq:RespLinearSystem}. The main advantage of this approach is that all linear equations can be solved in parallel. Additionally, we solve these equations with the minimal residual method \cite{Paige1975}, which, based on Lanczos iterations, takes advantages of the sparseness of $H_{\text{BdG}}$, and thus only requires sparse matrix-vector multiplications. The total computational time has therefore a close to linear scaling in the problem dimensions. For many Lanczos iterations, a known risk is that roundoff errors may give numerical instabilities from loss of orthogonality \cite{Cullum2002}. We have verified by comparing with direct diagonalization of $H_{\text{BdG}}$ for rotation angles down to $\theta \approx 1.5^{\circ}$,  that all calculated expectation values are within the expected accuracy with no evidence of numerical instability.
		
		\noindent{\bf Solving for and analyzing the superconducting order.}		
			We solve the gap equations, Eqs.~\eqref{eq:selfcons} and \eqref{eq:linearized}, using the rational pole expansion of the Fermi operator, Eqs.~\eqref{eq:PFMEXP} and \eqref{eq:PFMResp}, with a maximal error less than $10^{-8}\%$. 
			More specifically, to solve the linear gap equation we first compute the static response matrix $ \Gamma^{rs}_{ij}(T) = - \partial \langle s_{ij} \rangle / \partial \Delta_{rs}$ at the fixed temperatures $\TC = 6.25 \times 2^n 10^{-5} t \approx 2^n$~K for $n=0,1,...6$, with a uniform $\KS \times \KS$ grid sampling of reciprocal space ($\KS=4$). Then the largest eigenvalues $\lambda_n (\TC) $ of $\Gamma$ were found using the eigenvalue solver PRIMME \cite{PRIMME}. Finally, for each eigenvalue, the coupling strength of $J_n(\TC) = 1/\lambda_n (\TC)$ ensures that Eq.~\eqref{eq:linearized} is satisfied. That is, at $J_n(\TC)$ the $n^{\text{th}}$ (subordinate) order has the critical temperature of $\TC$. In Fig.\ref{fig:02 - TceV}(a) we present the $(J_n(\TC),\TC)$ pairs for the fourth largest eigenvalues at the fixed $\TC$.
						
			The eigenvectors corresponding to each solution $(J_n(\TC),\TC)$ are symmetry-classified on the \moire{} length scale. For this, we construct the irreducible representation (Irrep) projection operators $\hat{P}_{I} = ( d_I/ |G| ) \sum_{g \in G} \chi^{*}_{I} (g) \hat{\Gamma}(g) $ of the point group $G=D_3$, where $I$ index the Irreps of dimension $d_I$ with group characters $\chi_{I}$ and where $\hat{\Gamma}(g)$ is the action of the symmetry operations $g$ on all the superconducting order parameters in the \moire{} unit cell. 					
			Because an eigenvector of a symmetric static response matrix necessarily belongs to only a single Irrep, only the corresponding projection operator leaves the eigenvector invariant, while all other Irrep projection operators annihilates it, thereby allowing the symmetry of the eigenvector to be easily identified. 
			Nonetheless, the same classification is also obtained with more insight by considering only the linear subspace consisting of the six order parameters on the nearest-neighbor bonds of two layer-aligned $A$-sites, since together these six order parameters form a closed set under the symmetry operations $\hat{\Gamma}(g)$ of the $D_3$ symmetry group and necessarily with the same symmetry as the whole eigenvector. In particular, this linear subspace is the direct sum of the $D_3$ Irreps: $A1$, $A2$ and two $E$, with Irrep basis vectors (up to normalization) obtained from the non-zero eigenvectors of the projection operators and listed in Table \ref{tab:Irreps}. We then easily classify the \moire-scale symmetry of each eigenvector by projecting each subset of bonds on these Irrep basis vectors.						
			\begin{table}
				\begin{tabular}{ l l }
						Irrep 		& Basis vectors \\
						\hline
						A1: 			& $( 1,  1,  1,  1,  1,  1 	 )                 					      $  \\
						A2: 			& $( 1,  1,  1,  -1,  -1,  -1)              					        $  \\
						E:  			&	$( 2,  -1,  -1,  0,  0,  0 )$, $ ( 0,  1,  -1,  0,  0,  0 ) $  \\
						E:				&	$( 0,  0,  0,  2,  -1,  -1 )$, $ ( 0,  0,  0,  0,  1,  -1 ) $
				\end{tabular}
				\caption{
					{ \bf Irreducible representation decomposition.} Basis vectors of the $D_3$ irreducible representation (Irrep) decomposition of the six dimensional linear subspace of the superconducting nearest-neighbor bond order parameters surrounding the aligned carbons, pierced by the twist rotation axis. The first (last) three dimensions are for the bond order parameters in the top (bottom) layer.
				}
				\label{tab:Irreps}
			\end{table}
			
			As previously noted, the \moire{} lattice model also has an approximate $D_6$ symmetry, besides the exact $D_3$ symmetry \cite{Zou2018}. We have verified that also the leading superconducting order in Fig.\;\ref{fig:02 - TceV}(a) has an approximate $D_6$ symmetry by calculating that it has a $99\%$ weight in the $E_2$ Irrep of $D_6$. Here the weight $\| \hat{P}_{I} \hat{\Delta} \|^2 / \| \hat{\Delta} \|^2$ in the $D_6$ Irrep $I$ is calculated from the projection operators with the $D_6$ geometric transformations $\hat{\Gamma}(g)$ that interchanges all bonds of TBG with a zeroth order interpolation around the nearest hexagon center to the $\AAReg$ region.
			
			We also extract, throughout the \moire{} unit cell, the local atomic-scale symmetry of the superconducting order. We do this by using the basis vectors of the Irreps of the graphene symmetry group $D_{6h}$. Expressed on all nearest neighbor bonds these are 
				$\bm{f}_{s}= (1,1,1) / \sqrt{3}$ 
			for the identity representation $A_{1g}$ ($s$-wave symmetry), 
				$\bm{f}_{d_{x^2-y^2}}= (2,-1,-1) / \sqrt{6}$ ($d_{x^2-y^2}$-wave) and $\bm{f}_{d_{xy}} = (0,-1,1) / \sqrt{2}$ ($d_{xy}$-wave)
			for the two-dimensional $E_{2g}$ representation \cite{BlackSchaffer2007}.
			The atomic-scale symmetry of the three-dimensional bond order parameter vector 
				$
					\bm{\Delta}(\bm{x}_i) = (\Delta_{i j_1}, \Delta_{i j_2}, \Delta_{i j_3})
				$
			at site $\bm{x}_i$ is then extracted by projection on to these form factors.
			
			Complimentary to the linear equations  Eq.~\eqref{eq:linearized}, we also solve the fully non-linear self-consistency equations Eq.~\eqref{eq:selfcons} using Eq.~\eqref{eq:ExpLinearSystem} at an effective zero temperature ($T=10^{-7}t$)  and with the same k-point sampling ($\KS=4$) as the linear equation Eq.~\eqref{eq:linearized}. We find that the zero temperature self-consistent solutions $\bm{\hat{\Delta}}_{\rm SC}$ are always completely contained in the eigenspace $\Lambda_1$ of the leading solutions of the linear equation. Formally, this is established by the projection $\|P_{\Lambda_1} \bm{\hat{\Delta}}_{\rm SC}\|^2/\| \bm{\hat{\Delta}}_{\rm SC}\|^2$ being as high as $0.99993$ for $J=\WeakJ$ and $0.99992$ for $J=\StrongJ$, where $ \| \bm{\hat{\Delta}} \| = \sqrt{\sum_{ \langle i, j \rangle } |\Delta_{i j}|^2 }$ is the order parameter field norm.		
								
		\noindent{\bf Additional physical properties.}					
			We additionally compute a number of different physical properties using the following methods.
				\begin{description}[labelindent=\parindent,labelsep=\parindent,leftmargin=0.0em]
				\setlength\itemsep{-0.3em}
				\item[Band plot] For the band plot in Fig.~1(c), the $16$ lowest eigenvalues of the the normal state were computed evenly along the high symmetry cuts shown in Fig.~1(a) with PRIMME \cite{PRIMME}.
		
				\item[Density of states] 
				From the lowest eigenvalues computed with PRIMME, the density of states, $\text{DOS}(E) = \sum_{n} \delta(E-E_n)$, was computed as the kernel density estimation with Gaussian width $K_\sigma$, using on a regular $\KS \times \KS$ grid in reciprocal space for both the normal and the superconducting states, see Table \ref{tab:DOS} for parameters of each figure. Furthermore, in Figs.~1(d),(e), the DOS was calculated for the commensurate angles parametrized as $(p,q)$ with $p=1$ and $q=21,23,...,71$ and then normalized by the area of the \moire{} unit cell. For $q<31$ we used $K_\sigma=1$ meV and $\KS=32$, while for $31\leq q$, $K_\sigma=1/2$ meV and $\KS=16$. The energy of the VHSs shown in Fig.\ref{fig:01 - Normal}(d) were  defined as the position of the two local maximum in the resulting density estimation. Similarly, the value at the two local maxima define the VHS DOS in Fig.~\ref{fig:01 - Normal}(e). For Fig.~4(a), we used a two-dimensional Gaussian Kernel density estimation with a kernel widths of $1/10$ meV for the energy and $\pi/10$ for the interlayer phase.
				\begin{table}
					\begin{tabular}{ l l c c }
								Figure(s) 					& Spectrum 	& 	$\KS$ 				& $K_\sigma$ (meV) \\
								\hline
								Fig.~1(c) 					& Normal 		& 	$60$		 			& $1/20$           \\
								Fig.~1(d),(e) 			& Normal 		& 	$16$ or $32$	& $1/2$ or $1$     \\
								Fig.~2(h)-(j)				& BdG			 	& 	$48$ 					& $1/5$            \\
								Fig.~4(c)-(e) 			& BdG 			& 	$24$ 					& $2/5$
						\end{tabular}						
						\caption{
							{\bf Attributes of the density of states (DOS) figures.} The figure(s) in each row (first column) either show the DOS of the normal state or a superconducting Bogoliubov de-Gennes (BdG) Hamiltonian matrix spectrum (second column). In all figures the DOS is plotted as a kernel density estimation from a $\KS \times \KS$ point sampling of the Brillouin (third column) and with a Gaussian width of $K_\sigma$ (last column).
						}
						\label{tab:DOS}
				\end{table}
				
				\item[Local density of states] 
				From the lowest eigenvalues together with eigenvectors computed with PRIMME, the normal state local density of states (LDOS) of Fig.~1(g) was computed using 
				\begin{equation}
					n(\bm{x}_i,E) = \NK^{-1} \sum_{\lambda,\bm{k}} |U_{\lambda i}(\bm{k})|^2 \delta(E - E_{\bm{k} \lambda})
				\end{equation}
				where $U_{\lambda i}(\bm{k})$ is the $i$-site amplitude of the $\lambda$ eigenvector. 
				For the superconducting state in Fig.~3 we instead used the electronic part of the LDOS: 
					\begin{equation}
						n (\bm{x}_i,E) 
						= 
						\NK^{-1} \sum_{\lambda,\bm{k}}
							|u_{i\lambda}(\bm{k})|^2 \delta(E - E_{\bm{k} \lambda}) +
							|v_{i\lambda}(\bm{k})|^2 \delta(E + E_{\bm{k} \lambda})
					\end{equation}
				where $u_{i\lambda}(\bm{k})$ and $v_{i\lambda}(\bm{k})$ are the particle and hole amplitudes of the	Nambu-spinor eigenvector. For both cases, we computed the LDOS as the weighted Kernel density estimation with the same Gaussian kernel width $K_\sigma = 1/8$ meV and $\KS=24$.				
				
				\item[Charge density]
				The charge density of the \moire{} flat band $N_{\rm m}(\bm{x})$ in Fig.~1(f) was obtained by first computing the particle occupation $N_i(\mu) =\sum_{\sigma} \langle c^\dagger_{i\sigma} c_{i\sigma} \rangle$ using Eq.~\eqref{eq:PFMEXP} at each site $i$ of the top graphene layer for two different uniform chemical potentials: $\mu_1 = 7.5$ meV and $\mu_2 = 21.4$ meV. The lower (upper) $\mu$ is in the band gaps below (above) the \moire{} bands. The change in particle occupation between the two $\mu$s, $N_{\rm m} (\bm{x}_i) = N_i(\mu_2) - N_i(\mu_1)$, is thus the charge density of the \moire{} bands and is also equivalent to the integrated LDOS: $N_m (\bm{x}) = \int_{\mu_1}^{\mu_2} n(\bm{x},E) \dd E$. To obtain the joint charge density of both A and B sublattices with in each unit cell, the density on each sublattice was first linearly interpolated as a function of position and then summed on an hexagonal grid in Fig.~1(f).
				
				\item[Relative energy differences]			
				The relative energy differences in Figs.~2(e),(f), were calculated by reinserting the linear combinations $\bm{\hat{\Delta}}(\Theta,\varphi)$ of the $T=1.25 \times 10^{-4}t$ eigenvectors $\bm{\hat{\Delta}}_{x,y}$ in to the BdG Hamiltonian with same field amplitude 	$\| \bm{\hat{\Delta}} \|$ as the zero temperature self-consistent solution of the corresponding coupling strength $J(\TC)\approx \WeakJ$. The zero temperature energy difference between the $\bm{\hat{\Delta}}(\Theta,\varphi)$ state and the ground state at $\bm{\hat{\Delta}}(0,0)$ were then computed as 
					\begin{equation}
						\Delta E(\Theta,\varphi) 
						= 
						\sum_{n,\bm{k}} \left[ \theta( - E^{n}_{\bm{k}, \Theta, \varphi} ) E^{n}_{\bm{k}, \Theta, \varphi}  - \theta( -E^{n}_{\bm{k}, 0, 0} ) E^{n}_{\bm{k}, 0, 0} \right] 
					\end{equation}
				for the eigenenergy bands $E^{n}_{\bm{k}, \theta, \varphi}$ indexed by $n$. Note that all other terms in the free energy cancel between the two solutions. For the $k$-vector sum, the same $k$-point sampling ($\KS=4$) was used as when solving the gap equations.
				
				For the results in Fig.~2(g) we calculated in the same way the relative energy differences for $\bm{\hat{\Delta}}(\Theta,\varphi)$ but over a range of field norms $\| \bm{\hat{\Delta}} \|$, i.e.~not constrained to the self-consistent values at the appropriate coupling strength. Since the used basis vectors $\bm{\hat{\Delta}}_{x,y}(\TC)$ might change with $\TC$ and thus field norm, this is technically an approximation. However, we find that basis vectors $\bm{\hat{\Delta}}_{x,y}(\TC)$ of the leading solution are remarkably stable, and that the resulting energy differences are insensitive to the $\bm{\hat{\Delta}}_{x,y}(\TC)$ used for the variation, a result we checked by using both eigenvectors from $T=10^{-3}t $ and $T=1.25 \times 10^{-4}t$, as well as a higher ($\KS=6$) $k$-point sampling. To accurately capture the energy difference also in the very small $\| \bm{\hat{\Delta}} \|$ regime, we computed the energy difference using a very high $k$-point sampling ($\KS=20$). 								
				
				\item[Condensation energies] 
				The zero temperature condensation energies in Fig.~2(h-j), 
				as well as the the maximum condensation energy of the self-consistent $\phi=\pi$ state in Fig.~4(b),
				were computed using all the eigenvalues of the normal state and of the BdG Hamiltonian attained with Armadillo \cite{Sanderson2016,Sanderson2018}. In terms of the eigenvalues $E^n$ of Hamiltonian, the normal state energy is 
					$E_{\rm N} = \NK^{-1} \sum_{n,\bm{k}} \theta( -E^{n}_{\bm{k}} ) E^{n}_{\bm{k}}$, 
				while the energy in the superconducting state
					$
						E(\bm{\hat{\Delta}}) 
						= 
						\NK^{-1} \sum_{n,\bm{k}} \theta( -E_{\bm{k}}^{n} (\bm{\hat{\Delta}}) ) E_{\bm{k}}^{n} (\bm{\hat{\Delta}})
						- 2 \mu \NC 
						+ \sum_{\langle i,j \rangle } \left| \Delta_{ij}|^2 \right / J
					$,
				also contains the constant energy shift $C$ of the BdG form, see~Eq.~\eqref{Eq:BdGForm}. The zero temperature condensation energy is then just the difference $E_{\rm cond} = E_{\rm N} - E(\bm{\hat{\Delta}})$.				
				
				\item[Streamlines]			
				The streamlines in Figs.~2(b),(c) were constructed by numerically integrating the flow equation 
					$ \partial_{t} \bm{r}(t) = \bm{\chi} ( \bm{r} ) $ both forwards and backwards
					starting from initial points on a hexagonal grid throughout the \moire{} cell.	
					Here the vector field $\bm{\chi}(\bm{x}_i) = \cos \tau(\bm{x}_i) \bm{\hat{x}} + \sin \tau(\bm{x}_i) \bm{\hat{y}}$
					was the linearly interpolated from the angle
						$ 
							\tau( \bm{x}_i ) = 
							\arctan 
							( 
								\bm{\Delta}(\bm{x}_i) \cdot \bm{f}_{d_{x^2-y^2}}
								/ 
								\bm{\Delta}(\bm{x}_i)\cdot \bm{f}_{d_{xy} } 
							) 
						$, with separate accounting for the appropriate quadrant for the angle. 
			\end{description}

\section{Data Availability}
	Data are available from the corresponding authors upon request.

\section{Code Availability}
	Computer program source codes used for all calculations and analysis are available for review from the corresponding authors upon request.

\section{Acknowledgment}
	 We acknowledge financial support from the Swedish Research Council (Vetenskapsr\aa det Grant No.~2018-03488) and the Knut and Alice Wallenberg Foundation through the Wallenberg Academy Fellows program. The computations were enabled by resources in project SNIC 2020/5-18 provided by the Swedish National Infrastructure for Computing (SNIC) at UPPMAX, partially funded by the Swedish Research Council through grant agreement No.~2018-05973.
		
\section{Author contributions}
	A.B.S. and T.L. initiated the project. T.L. developed the computational framework and carried out the simulations with help from J.S. and F.P.. All authors analyzed and interpreted the results. A.B.S. and T.L. wrote the manuscript with input from all authors.

\section{Competing interests}
	The authors declare no competing interests.	 

%

\end{document}